\documentclass[preprint,aps,showpacs]{revtex4}

\usepackage{graphicx}
\usepackage{bm}       
\usepackage{mathptmx} 

\begin{document}

\title{Positivity and topology in lattice gauge theory}
\author{Michael Creutz}
\affiliation{
Physics Department, Brookhaven National Laboratory\\
Upton, NY 11973, USA
}

\begin{abstract}
{The admissibility condition usually used to define
the topological charge in lattice gauge theory is incompatible with
a positive transfer matrix.  }
\end{abstract}

\maketitle

With certain smoothness assumptions, continuum Yang Mills field
configurations in four dimensional space time can be classified by a
topological winding number \cite{thooft}.  This realization has played
a major role in our understanding of the importance of
non-perturbative phenomena in the $SU(3)$ gauge theory of the strong
interactions \cite{coleman}.

This winding number is uniquely defined for smooth fields; however,
for a quantum field theory one must integrate over all configurations,
some of which may not be sufficiently smooth for a unique definition
of the topological charge.  Regulating the theory on a lattice brings
in questions of how to handle these topological objects as their size
drops below the lattice spacing.  Considerable recent progress in this
area has involved the use of Dirac operators with exact symmetries
under chiral transformations \cite{gw,Luscher:1999mt,overlap}.
Indeed, a rigorous lattice extension of the continuum index theorem
relates the winding number to the zero eigenvalues of these chiral
operators.

Classifying fields by their winding number divides the space of
configurations into distinct topological sectors.  With conventional
actions, however, the configuration space is simply connected.  Thus
the winding number must be singular as one moves from one sector to
another \cite{Creutz:2002qa}.  The locations of these singularities
will in general depend on the particular Dirac operator used to define
the topology.  This ambiguity can be avoided by placing a constraint
on the roughness of the gauge fields
\cite{Luscher:1981zq,Hernandez:1998et,Neuberger:1999pz}.  As usually
formulated, the constraint forbids plaquettes to stray further from
the identity than a given distance.

At first sight this constraint seems quite harmless, and, indeed, it
is irrelevant to all perturbative physics.  However, in this note I
show that such a constraint is incompatible with requiring a positive
transfer matrix
\cite{Luscher:1976ms,Osterwalder:1977pc,Creutz:1976ch}.  The argument
builds on an old discussion of Grosse and Kuhnelt \cite{Grosse:1981vj}
that shows the failure of positivity for the Manton action
\cite{Manton:1980ts}.

I work with the gauge fields alone, and restrict myself to single
plaquette actions.  In the path integral, I assume the action
associates a real non-negative weight $W(P)$ to any given plaquette,
where the plaquette variable $P$ is in the gauge group.  I assume that
$W(P)$ is smooth, indeed analytic, for $P$ in some small vicinity of
the identity.  This insures a smooth mapping onto the perturbative
limit.  Away from the identity, I only assume it is piecewise smooth.
The admissibility condition states that $W(P)$ should vanish for $P$
in a finite region of the group some distance away from the identity.

To show that such a condition conflicts with positivity, I start by
paralleling the argument of Ref.~\cite{Grosse:1981vj} and reduce the
issue to a single time-like plaquette.  If positivity holds, the
matrix element of the transfer matrix between arbitrary states must be
non-negative.  In particular, for any square integrable function
$\psi(g)$ over the group, one must have
\begin{equation}
\int dg^\prime\ dg\ \psi^*(g^\prime) W({g^\prime}^{-1} g) \psi(g)
\ge 0.
\label{basic}
\end{equation}
As a violation of this for any subgroup would imply a violation for
the full group, I restrict the discussion to a $U(1)$ subgroup.
Denote the elements of this subgroup as $e^{i\theta}$, with
$\theta=0$ representing the identity element. I then should have
\begin{equation}
\int_{-\pi}^\pi d\theta^\prime\ d\theta\ \psi^*(\theta^\prime) 
W(e^{i(\theta-\theta^\prime)}) \psi(\theta)
\ge 0.
\end{equation}
for arbitrary square integrable $\psi(\theta)$.  Note that the
restriction to a $U(1)$ subgroup does not place any serious constraint
on the allowed values of the plaquette.  For example, in $SU(3)$ the
eighth Gell-Mann matrix generates a subgroup where ${1\over 3}{\rm
ReTr} g$ runs over the full allowed region from $-{1\over 2}$ to 1.

Reduced to a $U(1)$ subgroup, it becomes convenient to work with the
Fourier functions $\psi(\theta)=e^{in\theta}$.  Inserting one of these
into the above and changing variables to $\phi=\theta-\theta^\prime$
gives
\begin{equation}
f_n\equiv \int d\phi
W(e^{i\phi}) e^{in\phi}
\ge 0.
\end{equation}
Thus all Fourier components of $W$ must be real and non-negative, an
extremely strong constraint.  As is well known, any piecewise smooth
weight can be reconstructed from its Fourier components
\begin{equation}
W(e^{i\phi})={1\over 2\pi}\sum_{n=-\infty}^\infty f_n e^{-in\phi}. 
\end{equation}
Reality of the weight gives $f_{-n}=f_n$.  

I now extend $W$ into the complex plane.  For this I define
$z=e^{-i\phi}$, so that the physical weight function occurs on the
unit circle.  Separating the positive and negative terms in the series
with the definition
\begin{equation}
f_+(z)=\sum_{n=1}^\infty f_n\ z^n.
\end{equation}
I write
\begin{equation}
W(z)=f_0+f_+(z)+f_+(1/z).
\end{equation}

The assumption that the weight is analytic near $z=1$ coupled with the
positive nature of the $f_n$ implies that one can also expand $f_+(z)$
about the origin with a radius of convergence $z_0$ greater than
unity.  Thus, the function $f_+(z)$ is analytic inside a circle of
this radius about the origin.  For the remaining piece contained in
$f_+(1/z)$, I instead have an analytic function of $z$ outside a
circle of radius $1/z_0<1$.  Thus the full weight $W(z)$ must be an
analytic function in the common region, {\it i.e.} a ring with $1/z_0
< |z| <z_0$.
 
This analyticity immediately precludes many possible actions.  For the
present case, if $W(z)$ vanishes on any finite region of the unit
circle, it must vanish everywhere, contradicting using it as a weight
in a path integral.  This is the main result of this note.

Note that the weight can vanish at a finite number of discrete points.
For example $W=1+\cos(\theta)$ satisfies the positivity condition
while being zero at $\theta=\pi$.  It is only vanishing over a
continuous region that is forbidden.  The above proof also gives an
explicit procedure for finding a wave function for which the transfer
matrix is ill behaved; just calculate the Fourier coefficients
successively until you find one that is not positive.

The positivity of the Fourier coefficients is a special case of the
requirement that in a character expansion of $W(P)$, all coefficients
must be positive \cite{Bitar:1983tn}.  This follows from using
representation matrix elements for the wave functions in
Eq.~\ref{basic}.  This shows that the character expansion is
absolutely convergent, and the analyticity extends to the entire
group.  Except for possible isolated points, there must be a finite
probability of reaching any plaquette value.

So far the admissibility condition is the only way proven to give a
uniquely defined topological index.  However, this does not
necessarily preclude the existence of some other smoothness condition
to accomplish the same.  The generality of the present result shows
that any such condition can not be a local constraint depending only
on individual plaquettes.

Of course, positivity may not be a necessary requirement if the
non-positive effects disappear in the continuum limit.  Indeed, such
possibilities have been discussed in the context of generalized gauge
actions, {\it e.g.} Ref.~\cite{Luscher:1984is,Necco:2003vh}.  But it
seems a large price to pay just to define an esoteric object such as
the topological susceptibility.

As for the existence of the continuum Yang-Mills theory, it does not
appear that a non-perturbative ambiguity in the definition of the
topological susceptibility causes any harm.  This concept is rather
abstract, and it is not clear if it can be measured in any physical
experiment, even considering external sources.

With several species of degenerate quarks, there is one point where
the topological susceptibility is well defined.  This is the chiral
point, where the existence of massless Goldstone bosons uniquely fixes
the quark masses to zero.  Using a Ginsparg-Wilson formulation for the
fermions then ensures that the topological susceptibility vanishes.

For one flavor of massless quark, the issue is less clear.
Ref.~\cite{Creutz:2004fi} argues that in the one quark case, the
massless quark theory may have a scheme dependent continuum limit.  If
so, the point of vanishing topological susceptibility is also
ambiguous.

Going to the pure glue theory, {\it i.e.} $n_f=0$, the absence of a
fermion determinant will allow the gauge fields to become even
rougher.  Recent discussions \cite{Giusti:2004qd,Luscher:2004fu} of
measuring the topological susceptibility with an external
Ginsparg-Wilson operator have shown that all perturbative divergences
are controlled.  However, non-perturbatively, different operators have
the potential to give different answers for the susceptibility for the
same physical continuum limit.  This is ruled out if the admissibility
condition is satisfied, a condition inconsistent with positivity in
the regulated theory.

\section*{Acknowledgements}
This manuscript has been authored under contract number
DE-AC02-98CH10886 with the U.S.~Department of Energy.  Accordingly,
the U.S. Government retains a non-exclusive, royalty-free license to
publish or reproduce the published form of this contribution, or allow
others to do so, for U.S.~Government purposes.


\begin{thebibliography}{99}

\bibitem{thooft}
G.~'t Hooft,
Phys.\ Rev.\ D {\bf 14}, 3432 (1976)
[Erratum-ibid.\ D {\bf 18}, 2199 (1978)].

\bibitem{coleman}
S.~R.~Coleman,
in {\it C77-07-23.7}
HUTP-78/A004
{\it Lecture delivered at 1977 Int. School of 
Subnuclear Physics, Erice, Italy, Jul 23-Aug 10, 1977},
reprinted in S. Coleman, {\it Aspects of Symmetry,} 
(Cambridge University Press, 1988).

\bibitem{gw}
P.~H.~Ginsparg and K.~G.~Wilson,
Phys.\ Rev.\ D {\bf 25}, 2649 (1982).

\bibitem{Luscher:1999mt}
M.~Luscher,
Nucl.\ Phys.\ Proc.\ Suppl.\  {\bf 83}, 34 (2000)
[arXiv:hep-lat/9909150].

\bibitem{overlap}  H. Neuberger, Phys.~Lett. { B417} 
(1998) 141;
H. Neuberger, Phys.~Lett. { B427} (1998) 353;
R. Narayanan and H.~Neuberger, 
Phys.~Lett.~B302 (1993) 62;
Phys.~Rev.~Lett.~71 (1993) 3251; 
Nucl.~Phys.~B412 (1994) 574;
Nucl.~Phys.~B443 (1995) 305.

\bibitem{Creutz:2002qa}
M.~Creutz,
Nucl.\ Phys.\ Proc.\ Suppl.\  {\bf 119}, 837 (2003)
[arXiv:hep-lat/0208026].

\bibitem{Luscher:1981zq}
M.~Luscher,
Commun.\ Math.\ Phys.\  {\bf 85}, 39 (1982).

\bibitem{Hernandez:1998et}
P.~Hernandez, K.~Jansen and M.~Luscher,
Nucl.\ Phys.\ B {\bf 552}, 363 (1999)
[arXiv:hep-lat/9808010].

\bibitem{Neuberger:1999pz}
H.~Neuberger,
Phys.\ Rev.\ D {\bf 61}, 085015 (2000)
[arXiv:hep-lat/9911004].

\bibitem{Luscher:1976ms}
M.~Luscher,
Commun.\ Math.\ Phys.\  {\bf 54}, 283 (1977).

\bibitem{Osterwalder:1977pc}
K.~Osterwalder and E.~Seiler,
Annals Phys.\  {\bf 110}, 440 (1978).

\bibitem{Creutz:1976ch}
M.~Creutz,
Phys.\ Rev.\ D {\bf 15}, 1128 (1977).


\bibitem{Grosse:1981vj}
H.~Grosse and H.~Kuhnelt,
Nucl.\ Phys.\ B {\bf 205}, 273 (1982).

\bibitem{Manton:1980ts}
N.~S.~Manton,
Phys.\ Lett.\ B {\bf 96}, 328 (1980).

\bibitem{Bitar:1983tn}
K.~M.~Bitar,
Max Planck Institute, Munich, report MPI-PAE/PTh 68/83 (unpublished).

\bibitem{Luscher:1984is}
M.~Luscher and P.~Weisz,
Nucl.\ Phys.\ B {\bf 240}, 349 (1984).

\bibitem{Necco:2003vh}
S.~Necco,
Nucl.\ Phys.\ B {\bf 683}, 137 (2004)
[arXiv:hep-lat/0309017];
arXiv:hep-lat/0306005 (PhD Thesis).

\bibitem{Creutz:2004fi}
M.~Creutz,
Phys.\ Rev.\ Lett.\  {\bf 92}, 162003 (2004)
[arXiv:hep-ph/0312225].

\bibitem{Giusti:2004qd}
L.~Giusti, G.~C.~Rossi and M.~Testa,
Phys.\ Lett.\ B {\bf 587}, 157 (2004)
[arXiv:hep-lat/0402027].

\bibitem{Luscher:2004fu}
M.~Luscher,
arXiv:hep-th/0404034 (unpublished).

\end{thebibliography}
\end{document}